\documentclass[10pt,twocolumn,letterpaper]{article}

\usepackage{iccv}
\usepackage{times}
\usepackage{epsfig}
\usepackage{graphicx}
\usepackage{amsmath}
\usepackage{amssymb}
\usepackage{multirow}
\usepackage{color}
\graphicspath{}

\usepackage[breaklinks=true,bookmarks=false]{hyperref}

\iccvfinalcopy 

\def\iccvPaperID{34} 

\ificcvfinal\fi

\begin{document}

\title{Adaptive Densely Connected Single Image Super-Resolution}

\author{Tangxin Xie, *Xin Yang, Chen Zhu, Xiaochuan Li\\
Department of Automation, Nanjing University of Aeronautics and Astronautics\\
Nanjing, China\\
{\tt\small yangxin@nuaa.edu.cn}
\and
Yu Jia\\
China Electronic Science and Technology Network Information Security Co., Ltd.\\
Chengdu, China\\
}

\maketitle
\ificcvfinal\thispagestyle{empty}\fi

\begin{abstract}
   For a better performance in single image super-resolution(SISR), we present an image super-resolution algorithm based on adaptive dense connection (ADCSR). The algorithm is divided into two parts: BODY and SKIP. BODY improves the utilization of convolution features through adaptive dense connections. Also, we develop an adaptive sub-pixel reconstruction layer (AFSL) to reconstruct the features of the BODY output. We pre-trained SKIP to make BODY focus on high-frequency feature learning. The comparison of PSNR, SSIM, and visual effects verify the superiority of our method to the state-of-the-art algorithms.
\end{abstract}

\section{Introduction}

Single image super-resolution aims at reconstructing an accurate high-resolution image from the low-resolution image. Since deep learning made big progress in the computer version, many SISR algorithms based on deep Convolution Neural Networks (CNN) have been proposed in recent years. The powerful feature representation and end-to-end training skill of CNN makes a huge breakthrough in SISR.

Dong \etal ~\cite{dong2014learning} first proposed SRCNN by introducing a three-layer CNN for image SR. Kim et al. increased the number of layers to 20 in VDSR~\cite{kim2016accurate} and DRCN~\cite{kim2016deeply},making notable improvements over SRCNN. As we all know, the deeper the network is, the more powerful the representation it has. However, with the depth of network grow, gradient disappear and gradient explosion will be the main problem to hinder the performance of the network. This problem was solved when He \etal ~\cite{he2016deep} proposed residual net (ResNet), and Huang \etal ~\cite{Huang_2017_CVPR} proposed dense net (DesNet). Many large scale networks were introduced in SISR, such as SRResNet~\cite{ledig2017photo}, EDSR~\cite{lim2017enhanced}, SRDenseNet~\cite{tong2017image}, RDN~\cite{zhang2018residual} etc.These methods aim at building a deeper network to increase the performance. Other methods such as RCAN~\cite{zhang2018image} and SAN~\cite{dai2019second} try to learn the correlation of the features in the middle layers. 

WDSR~\cite{yu2018wide} allows for better network performance with less computational effort. AWSRN~\cite{wang2019lightweight} applies an adaptive weighted network. Weight adaptation is achieved by multiplying the residual convolution and the residual hopping by coefficients respectively, and the coefficients can be trained. Since the performance of dense connections is better than the residual~\cite{lim2017enhanced}~\cite{zhang2018residual}, we develop an adaptive densely connection method to enhance the efficiency of feature learning. There is a similar global SKIP, a single sub-pixel convolution, in WDSR~\cite{yu2018wide}and AWSRN~\cite{wang2019lightweight}. Although the SKIP is set to recover low-order frequencies, there is no practical measure to limit its training. We present an adaptive densely connected super-resolution reconstruction algorithm (ADCSR). The algorithm is divided into two parts: BODY and SKIP. BODY is focused on high-frequency information reconstruction through pre-training the SKIP. ADCSR obtained the optimal SISR performance based on bicubic interpolation. There are three main tasks:

(1)WDSR~\cite{dai2019second}is optimized using adaptive dense connections. Experiments were carried out by initializing the adaptive parameters and optimizing the models. Based on the above efforts, the performance of the network has been greatly improved;

(2)We propose the AFSL model to perform image SR through adaptive sub-pixel convolution;

(3)We develop a method which pre-train SKIP first and then train the entire network at the same time. Thus, the BODY is focused on the reconstruction of high-frequency details to improve network performance.

\section{Related Works}

SISR has important applications in many fields, such as security and surveillance imaging~\cite{zou2011very}, medical imaging~\cite{shi2013cardiac}, and image generation~\cite{karras2017progressive}. The simplest method among them is the interpolation, such as linear interpolation, bicubic interpolation, and so on. This method takes the average of the pixel points in the known LR image as the missing pixel of the HR image. Interpolation works well in the smooth part of the image, but it works poorly in the edge regions, causing ringing and blurring. Additionally, learning-based and reconstruction-based methods are more complex such as sparse coding~\cite{yang2010image}, neighborhood embedded regression~\cite{chang2004super}~\cite{timofte2013anchored}, random forest~\cite{schulter2015fast}, etc.

Dong et al. first proposed a Convolutional Neural Network (CNN)-based super-resolution reconstruction network (SRCNN)~\cite{dong2014learning}, which performance is better than the most advanced algorithm at the time. Later, Shi et al. proposed a sub-pixel convolution super-resolution reconstruction network~\cite{shi2016real}. The network contains several convolutional layers to learn LR image features. Reconstruction is performed using the proposed sub-pixel convolutional layer. We can directly reconstruct the image utilizing the convolutional features from the deep convolutional network. Lim et al. proposed an enhanced depth residual network (EDSR)~\cite{lim2017enhanced}, which made a significant performance through the deeper network. Other deep network like RDN~\cite{zhang2018residual} and MemNet~\cite{tai2017memnet}, are based on dense blocks. Some networks focus on feature correlations in channel dimension, such as RCAN~\cite{zhang2018image}and SAN~\cite{dai2019second}. 

The WDSR~\cite{yu2018wide}proposed by Yu et al. draws two conclusions. First, when the parameters and calculations are the same, the model with more features before the activation function has better performance. Second, weight normalization (WN layer) can improve the accuracy of the network. In WDSR, there is a broader channel before the activation function of each residual block. Wang et al. proposed an adaptive weighted super-resolution network(AWSRN) based on WDSR~\cite{yu2018wide}. It designs a local fusion block for more efficient residual learning. Besides, an adaptive weighted multi-scale model is developed. The model is used to reconstruct features and has superior performance in methods with roughly equal parameters.

Cao et al. proposed an improved Deep Residual Network (IDRN)~\cite{cao2019fast}. It makes simple and effective modifications to the structure of residual blocks and skip-connections. Besides, a new energy-aware training loss EA-Loss was proposed. And it employs lightweight networks to achieve fast and accurate results. The SR feedback network (SRFBN)~\cite{li2019feedback} proposed by Li et al. applies the RNN with constraints to process feedback information and performs feature reuse.

The Deep Plug and Play SR Network (DPSR) ~\cite{zhang2019deep} proposed by Zhang et al. can process LR images with arbitrary fuzzy kernels. Zhang et al.~\cite{zhang2019zoom} obtained real sensor data by optical zoom for model training. Xu et al.~\cite{xu2019towards}generated training data by simulating the digital camera imaging process. Their experiments have shown that SR using raw data helps to restore fine detail and clear structure.
\begin{figure*}[htp]
	\centering
	\includegraphics[width=1\linewidth]{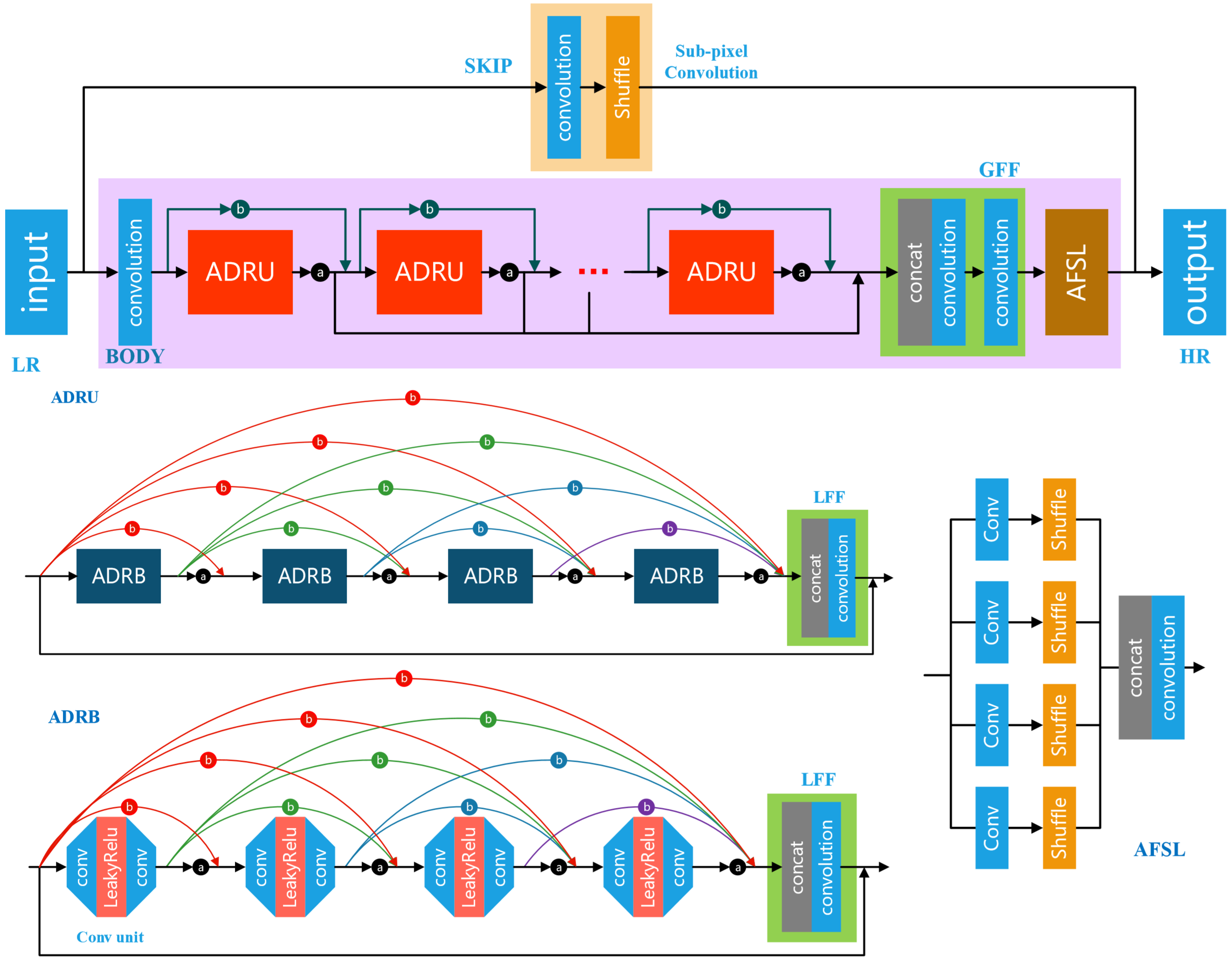}
	\caption{The architecture of our proposed Adaptive Densely Connected Super-Resolution Reconstruction (ADCSR).The top is ADCSR,the middle are ADRU and AFSL,the bottom is ADRB. }
	\label{fig:moedl}
\end{figure*}
\section{Our Model}
\subsection{Network Architecture}
As shown in Figure \ref{fig:moedl}, our ADCSR mainly consists two parts: SKIP and BODY. The SKIP just uses the sub-pixel convolution~\cite{shi2016real}. The BODY includes multiple ADRUs (adaptive, dense residual units), GFF (global feature fusion layer)~\cite{zhang2018residual}, and an AFSL layer(adaptive feature sub-pixel reconstruction layer). The model takes the RGB patches from the LR image as input. On the one hand, the HR image is reconstructed by SKIP using the low-frequency information of the LR images. On the other hand, the image is reconstructed by BODY using the high-frequency information of the LR images. We can obtain the final complete reconstructed HR image by combining the results of SKIP and BODY.

SKIP consists of a single or multiple sub-pixel convolutions with a convolution kernel size of 5. we have:
\begin{equation}
HR_{SKIP}=f_{sub-conv5}(I_{LR})\label{equo1}
\end{equation}
where $HR_{SKIP}$ represents the output of skip part, $I_{LR}$ denotes the input image of LR and $f_{sub-conv5}$ represents the sub-pixel convolution, which convolution kernel size is 5.

In the BODY, first, we use a convolution layer to extract the shallow features from the LR image.
\begin{equation}
F_{f}=f_{conv3}(I_{LR})\label{equo2}
\end{equation}
where $f_{conv3}$ represents the feature extraction convolution, which kernel size is 3.

Second, we use several ADRUs to extract the deep features. There are four ADRBs (adaptive dense residual blocks) through adaptive dense connections in Each ADRU. The features are merged by the LFF (Local Feature Fusion Layer) and combined with a skip connection as the output of the ADRU.  Each ADRB combines four convolution units by the same adaptive dense connection structure as ADRU. The convolution units adopt a convolution structure, which is similar to WDSR~\cite{yu2018wide}, including two layers of wide active convolution and one layer of Leakyrelu. After that, we fuse features by LFF, which combined with a skip connection as the output of the ADRB. GFF fuses the outputs of multiple ADRUs by means of concatenation and convolution.
\begin{equation}
X_{ADRUk+1}=b_{k}X_{ADRUk}+a_{k}f_{ADRUk}(X_{ADRUk})\label{equo3}
\end{equation}
where $X_{ADRUk}$ denotes the input feature map of $k$th ADRU, $f_{ADRUk}$ means the function of $k$th ADRU, $a_{k}$, $b_{k}$ are hyperparameters.
\begin{equation}
\begin{split}
Y_{ADRUk}=&a_{k}f_{ADRUk}(X_{ADRUk})\\
Y_{ADRUlast}=&b_{last}X_{ADRUlast}+\\
&a_{last}f_{ADRUlast}(X_{ADRUlast})
\end{split}
\label{equo4}
\end{equation}
$Y_{ADRUk}$ means the output of the $k$th ADRU, and $Y_{ADRUlast}$ represents the output of the last ADRU, which includes the skip connection.

The third part of BODY uses the GEF to combine all the output of ADRU, which fuses features by two convotion layers.
\begin{equation}
\begin{split}
F_{GFF}=f_{conv3}(f_{conv1}(&concat(Y_{ADRU1},\\
&...,Y_{ADRUn})))\label{equo5}
\end{split}
\end{equation}
where $concat$ means feature fusion.

Finaly, Image upsampling via AFSL. The AFSL consists of four sub-pixel convolution branches of different scales with a convolution kernel size of 3, 5, 7, and 9, respectively. The output is obtained through the junction layer and a single layer convolution.
\begin{equation}
\begin{split}
F_{AFSL}=f_{conv1}(&concat(f_{sub-conv3}(F_{GFF}),\\
&f_{sub-conv5}(F_{GFF}),\\
&f_{sub-conv7}(F_{GFF}),\\
&f_{sub-conv9}(F_{GFF})))\label{equo6}
\end{split}
\end{equation}

In the second stage of BODY, the feature amplification layer is also implemented by a single convolution layer. The whole BODY is:
\begin{equation}
\begin{split}
HR_{BODY}=F_{AFSL}((F_{GFF}(F_{f}(I_{LR}))))\label{equo7}
\end{split}
\end{equation}
$HR_{BODY}$represents the output of BODY. The whole network can be expressed by formulas \eqref{equo8}.
\begin{equation}
HR=HR_{BODY}+HR_{SKIP}\label{equo8}
\end{equation}
\subsection{ADRB and ADRU}
We will demonstrate the superiority of the adaptive dense connection structure in Chapter 4. To use as much adaptive residual structure as possible, we split the ADRU into ADRBs using adaptive dense connections and split ADRB into densely connected convolution units. At the same time, to get better results with less parameter amount, we use the residual block in WDSR as our convolution unit. As shown in Figure \ref{fig:moedl}, ADRB and ADRU have similar connection structure. ADRB contains four convolution units, each of which can be represented by equotion \eqref{equo9}.
\begin{equation}
f_{conv-unit}=f_{1conv3}(LeakyRelu(f_{2conv3}(x)))\label{equo9}
\end{equation}
where $x$ means the input of the convolution units. The kernel size of the  $f_{1conv3}$ is $[3,3,feats,3\times feats]$, and the $f_{2conv3}$ is $[3,3,3\times feats,feats]$, $feat$ is the input channels of the convolution units. 

The whole ADRB can be expressed by equation \eqref{equo10}.
\begin{equation}\label{equo10}
\begin{split}
&Y_{1}=f_{conv\_unit1}(x)\\
&X_{1}=a_{01}(Y_{1})+b_{01}(x)\\
&Y_{2}=f_{conv\_unit2}(X_{1})\\
&X_{2}=a_{12}(Y_{2})+b_{12}(Y_{1})+b_{02}(x)\\
&Y_{3}=f_{conv\_unit3}(X_{2})\\
&X_{3}=a_{23}(Y_{3})+b_{23}(Y_{2})+b_{13}(Y_{1})+b_{03}(x)\\
&Y_{4}=f_{conv\_unit4}(X_{3})\\
&\begin{split}
f_{ADRB}=f_{conv1}(&concat(a_{34}Y_{4},b_{34}Y_{3},\\
&b_{24}Y_{2},b_{14}Y_{1},b_{04}x))+x
\end{split}
\end{split}
\end{equation}
where $f_{conv\_unit1}$means convolution unit,  $x$ denotes the input of ADRB.  $a_{mn}, b_{mn}$ are hyperparameter, $X_i$ denotes the input of $(i+1)$th convolution unit, $Y_{j}$represents the output of $j$th convolution unit.

The whole ADRU can be formulated by equation \eqref{equo11}.
\begin{equation}\label{equo11}
\begin{split}
&Y_{1}=f_{ADRB1}(x)\\
&X_{1}=a_{01}(Y_{1})+b_{01}(x)\\
&Y_{2}=f_{ADRB2}(X_{1})\\
&X_{2}=a_{12}(Y_{2})+b_{12}(Y_{1})+b_{02}(x)\\
&Y_{3}=f_{ADRB3}(X_{2})\\
&X_{3}=a_{23}(Y_{3})+b_{23}(Y_{2})+b_{13}(Y_{1})+b_{03}(x)\\
&Y_{4}=f_{ADRB4}(X_{3})\\
&\begin{split}
f_{ADRU}=f_{conv1}(&concat(a_{34}Y_{4},b_{34}Y_{3},\\
&b_{24}Y_{2},b_{14}Y_{1},b_{04}x))+x
\end{split}
\end{split}
\end{equation}

\subsection{Implementation}
In this section, we will give specific implementation details. In SKIP, the convolution channel for the sub-pixel convolutional layer is defined as 5. The convolution kernel size of the LFF in BODY is 1.The two convolution kernel sizes of GFF are 1 and 3, respectively. In AFSL, the convolution kernels are 3, 5, 7 and 9. All other convolution kernel sizes are set to 3. There are 4 ADRUs in BODY. The number of output channels in feature extraction layer, convolution unit, LFF, and GFF are 128, and the 4 sub-pixel convolutions and the final output in AFSL are 3. The stride size is 1 throughout the network while using Leakyrelu as the activation function.

\section{Experiments}
\subsection{Adaptive dense connections}
\begin{figure}[htp]
	\centering
	\includegraphics[scale=0.3]{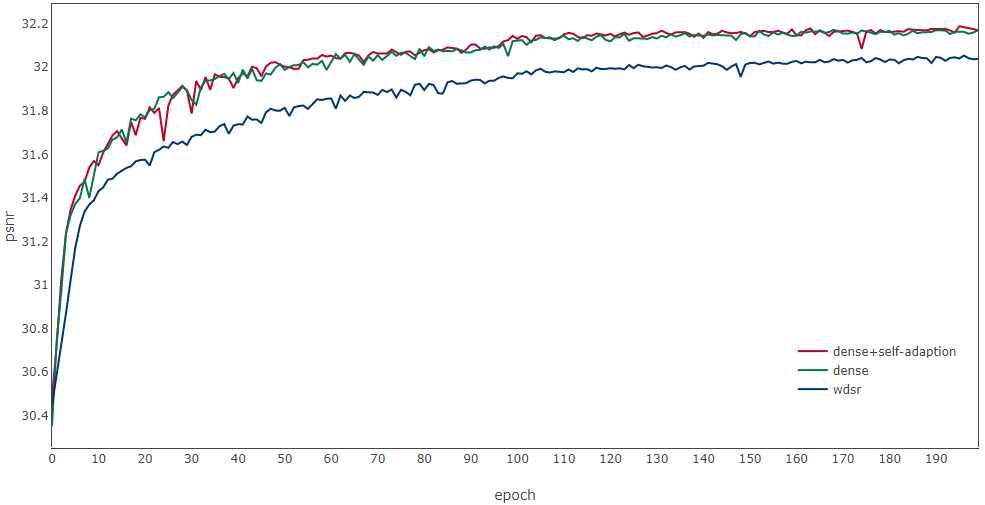}
	\caption{Convergence analysis of tests on B100 with scaling factor $\times2$ during different model structures}\label{fig-shuiping}
	\label{fig:adc}
\end{figure}

We propose a structure for adaptive dense connections such as ADRU, and verify its performance through experiments. In the experiment, we designed three models. The model parameters are the same, and the calculations are roughly equal. The structure of the models is similar to the ADCSR consisting of a single ADRU. These three models are: \\
a. Add LFF~\cite{zhang2018residual} on WDSR~\cite{dai2019second} (to obtain the same model depth);\\
b. Add a dense connection based on a;\\
c. Add parameter adaptation based on b.\\
The three models have the same training parameters. We train our models with the data set DIV2K~\cite{lim2017enhanced}. We also compare the performance on the standard benchmark dataset: B100~\cite{martin2001database}. The number of iterations is 200. The learning rate is $1\times10^{-4}$ and halved at every 100 epochs. As shown in Figure \ref{fig:adc}, networks with dense connections and parameter adaptation have the highest performance under the same conditions.

\subsection{Adaptive sub-pixel reconstruction layer (AFSL)}
\begin{figure}[htp]
	\centering
	\includegraphics[scale=0.3]{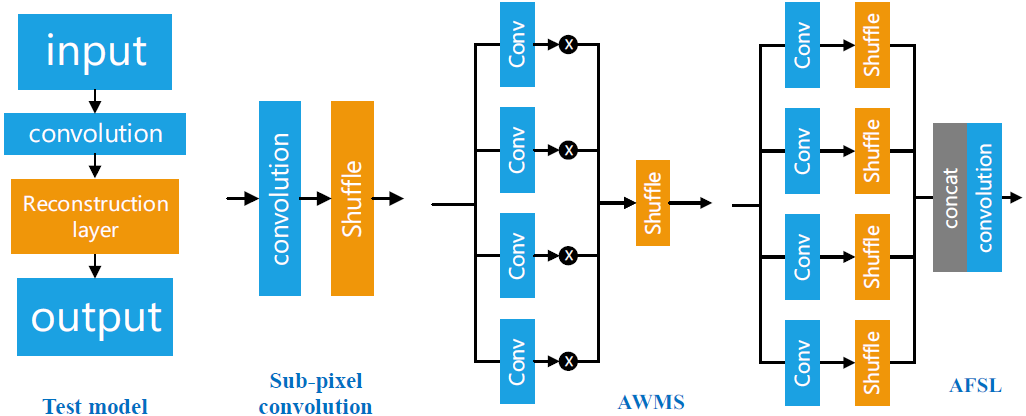}
	\caption{Test model and structural Comparison of Three Reconstruction Layers}\label{fig-shuiping}
	\label{fig:afsl}
\end{figure}

We test the reconstruction layer in BODY. We have designed a new reconstruction network model AFSL. To verify the performance of the model, we designed a straightforward model for comparison experiments. The model only includes the feature extraction layer and the reconstruction layer. As shown in Figure \ref{fig:afsl}, the reconstruction layers are Sub-pixel convolution~\cite{shi2016real}, AWMS~\cite{wang2019lightweight}, and AFSL. We performed the task on scale $\times2$. The feature extraction layers and experimental parameters of the models are the same. We tested the models with B100~\cite{martin2001database} and Urban100~\cite{huang2015single}. At the same time, we also analyzed the difference in the number of FLOPs and model parameters. The result is shown in Table \ref{tab:tafsl}. We can see that AWMS and AFSL require more calculations and parameters than Sub-pixel convolution while its performance is better. In the case where the setting and the calculated amount are the same, the performance of AFSL is slightly better than AWMS.

\begin{table}
	\centering
	\caption{Performance comparison of three reconstruction layers}
	\begin{tabular}[htp]{ccccc}
		\hline
		&B100&Urban100&FLOPs&Params\\		\hline
		Sub-conv&30.402&27.750&0.02G&9K\\
		AWMS&30.590&27.956&0.30G&128K\\
		AFSL&30.592&27.958&0.30G&128K\\\hline
	\end{tabular}
	\label{tab:tafsl}
\end{table}

\subsection{Pre-training SKIP}
We have explored a training method that performs a separate pre-training of SKIP while training the entire model. This training method is used to make SKIP focus on the reconstruction of low-frequency information, while BODY focuses on high-frequency information reconstruction.
We employ the same model, that is, the ADCSR containing a single ADRU with the same training parameters. But we train the model in different ways:\\
a. Train the entire network directly;\\
b. First pre-train SKIP, then train the whole network at the same time;\\
c. First pre-train SKIP, then set SKIP to be untrainable when training the entire network.

\begin{figure}[htp]
	\centering
	\includegraphics[scale=0.18]{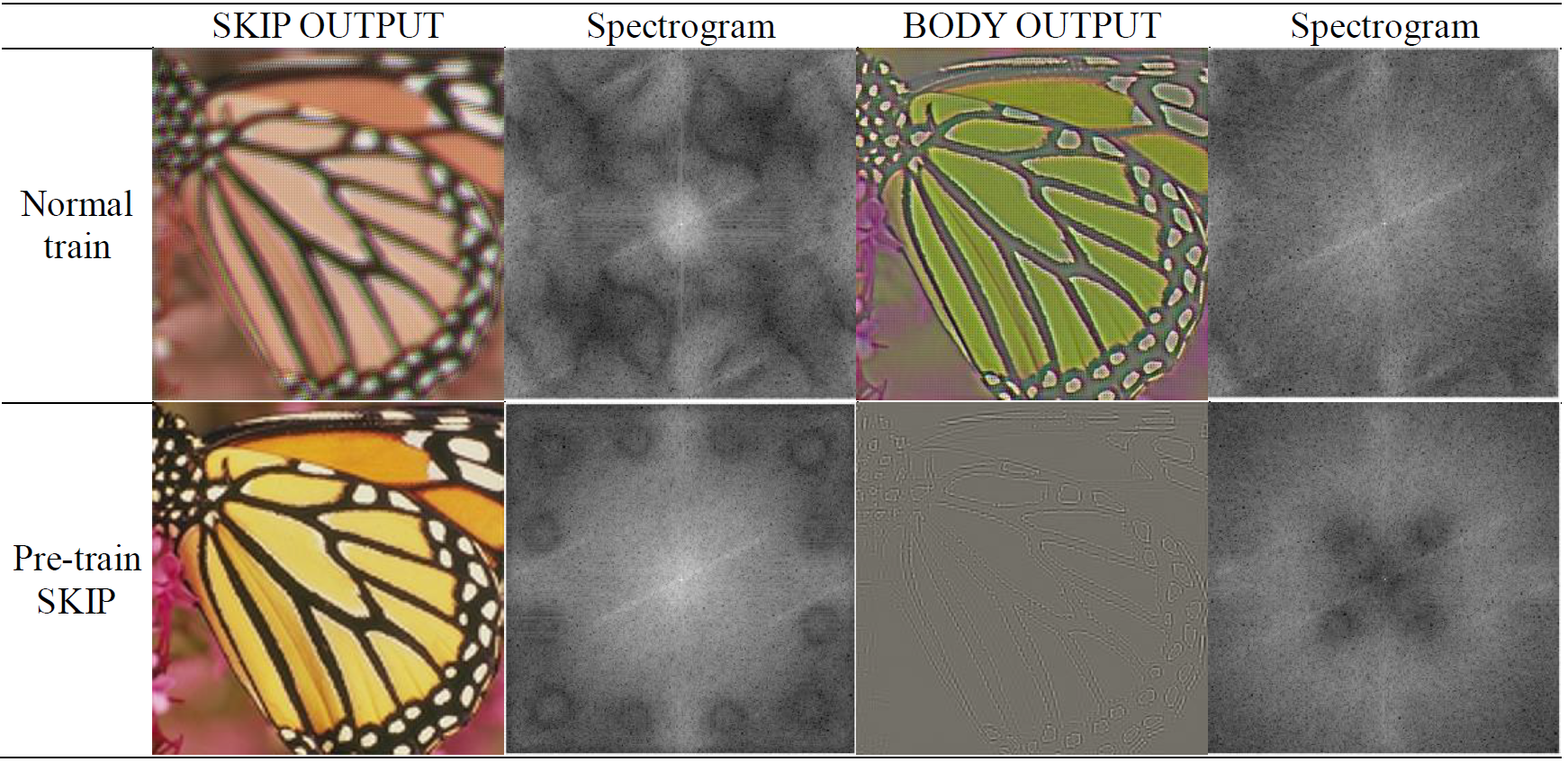}
	\caption{Results of pre-training SKIP on SKIP output}
	\label{fig:skipstop1}
\end{figure}

Figure \ref{fig:skipstop1} compares the image and image spectrum of SKIP and BODY output for models a and b. By comparing the output images, it can be seen that the BODY of the pre-trained SKIP model focuses on learning the texture edge details of the image. From the comparison of the output spectrum of the BODY part, the spectrogram of the pre-trained SKIP model is darker near the center and brighter around. It proves that the proposed method makes the BODY use more high-frequency information and less low-frequency information.

\begin{figure}[htp]
	\centering
	\includegraphics[scale=0.3]{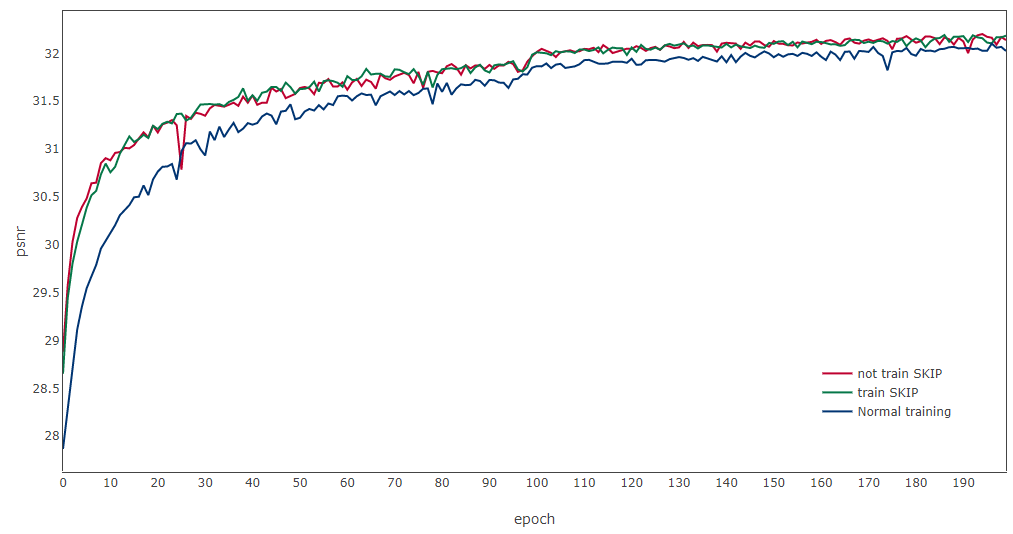}
	\caption{Convergence analysis of tests on B100 with scaling factor $\times2$ during different model training methods}
	\label{fig:skipstop2}
\end{figure}

Figure \ref{fig:skipstop2} is a comparison of the test curves of the model on the B100 under different training modes. We found that networks that were pre-trained with SKIP achieved higher performance. And the network performance of tests b and c are similar.

\subsection{Training settings}
We train our network with dataset DIV2K and Flickr2K~\cite{lim2017enhanced}. The training set has a total of 3,450 images without data augmentation. DIV2K is composed of 800 images for training while 100 images each for testing and validation. Flickr2K has 2,650 training images. The input image block size is $48\times48$. SKIP is trained separately, and then the entire network is trained at the same time. The initial learning rate is $1\times10^{-4}$. When the learning rate drops to $5\times10^{-7}$, the training stops. we also adopt L1 loss to optimize our model. We train the network of scale $\times2$ firstly. Subsequently, when training the network of scale  $\times3$, $\times4$, the BODY parameter of the scale $\times2$ is loaded (excluding the parameters of the AFSL). We train the model through the NVIDIA RTX2080Ti. Pytorch1.1.0+Cuda10.0+cudnn7.5.0 is selected as the deep learning environment.

\begin{table*}[htp]
	\begin{center}
		\caption{Quantitative evaluation of competing methods. We report the performance of state-of-the-art algorithms on widely used publicly available datasets, in terms of PSNR (in dB) and SSIM. The best results are highlighted with {\color{red}read} color while the {\color{blue}blue} color represents the second-best SR.}
		\label{tab:qecm}
		\begin{tabular}{|c|c|c|c|c|c|c|c|c|c|c|c|}
			\hline
			\multirow{2}{1.8cm}{method} & \multirow{2}{*}{scale} &\multicolumn{2}{|c|}{Set5~\cite{bevilacqua2012low}}&\multicolumn{2}{|c|}{Set14~\cite{yang2010image}}&\multicolumn{2}{|c|}{B100~\cite{martin2001database}}&\multicolumn{2}{|c|}{Urban100~\cite{huang2015single}} &\multicolumn{2}{|c|}{manga109~\cite{matsui2017sketch}} \\ \cline{3-12}
			& &PSNR &SSIM &PSNR &SSIM &PSNR &SSIM &PSNR &SSIM &PSNR &SSIM \\ 
			\hline\hline
			\multirow{11}{1.8cm}{Bicubic\newline SRCNN~\cite{dong2014learning}\newline VDSR~\cite{kim2016accurate}\newline LapSRN~\cite{lai2017deep}\newline MemNet~\cite{tai2017memnet}\newline EDSR~\cite{lim2017enhanced}\newline RDN~\cite{zhang2018residual}\newline RCAN~\cite{zhang2018image}\newline SAN~\cite{dai2019second}\newline ADCSR\newline ADCSR+}&\multirow{11}{*}{$\times2$}&33.66&0.9299&30.24&0.8688&29.56&0.8431&26.88&0.8403&30.80&0.9299 \\
			& &36.33&0.9542&32.45&0.9067&31.36&0.8879&29.50&0.8946&35.60&0.9663\\
			& &37.53&0.9590&33.05&0.9130&31.90&0.8960&30.77&0.9140&37.22&0.9750\\
			& &37.52&0.9591&33.08&0.9130&31.08&0.8950&30.41&0.9101&37.27&0.9740\\
			& &37.78&0.9597&33.28&0.9142&32.08&0.8978&31.31&0.9195&37.72&0.9740\\
			& &38.11&0.9602&33.92&0.9195&32.32&0.9013&32.93&0.9351&39.10&0.9773\\
			& &38.24&0.9614&34.01&0.9212&32.34&0.9017&32.89&0.9353&39.18&0.9780\\
			& &38.27&0.9614&34.12&0.9216&32.41&0.9027&33.34&0.9384&39.44&0.9786\\
			& &38.31&{\color{red}0.9620}&34.07&0.9213&32.42&0.9028&33.10&0.9370&39.32&0.9792\\
			&&{\color{blue}38.33}&{\color{blue}0.9619}&{\color{blue}34.48}&{\color{blue}0.9250}&{\color{blue}32.47}&{\color{blue}0.9033}&{\color{blue}33.61}&{\color{blue}0.9410}&{\color{blue}39.84}&{\color{blue}0.9798}\\
			&&{\color{red}38.38}&{\color{red}0.9620}&{\color{red}34.52}&{\color{red}0.9252}&{\color{red}32.50}&{\color{red}0.9036}&{\color{red}33.75}&{\color{red}0.9418}&{\color{red}39.97}&{\color{red}0.9800}\\
			\hline\hline
			\multirow{11}{1.8cm}{Bicubic\newline SRCNN~\cite{dong2014learning}\newline VDSR~\cite{kim2016accurate}\newline LapSRN~\cite{lai2017deep}\newline MemNet~\cite{tai2017memnet}\newline EDSR~\cite{lim2017enhanced}\newline RDN~\cite{zhang2018residual}\newline RCAN~\cite{zhang2018image}\newline SAN~\cite{dai2019second}\newline ADCSR\newline ADCSR+}&\multirow{11}{*}{$\times3$}&30.39&0.8682&27.55&0.7742&27.21&0.7385&24.46&0.7349&26.95&0.8556 \\
			& &32.75&0.9090&29.30&0.8215&28.41&0.7863&26.24&0.7989&30.48&0.9117\\
			& &33.67&0.9210&29.78&0.8320&28.83&0.7990&27.14&0.8290&32.01&0.9340\\
			& &33.82&0.9227&29.87&0.8320&28.82&0.7980&27.07&0.8280&32.21&0.9350\\
			& &34.09&0.9248&30.00&0.8350&28.96&0.8001&27.56&0.8376&32.51&0.9369\\
			& &34.65&0.9280&30.52&0.8462&29.25&0.8093&28.80&0.8653&34.17&0.9403\\
			& &34.71&0.9296&30.57&0.8468&29.26&0.8093&28.80&0.8653&34.13&0.9484\\
			& &34.74&0.9255&30.65&0.8482&29.32&0.8111&29.09&0.8702&34.44&0.9499\\
			& &34.75&0.9300&30.59&0.8476&29.33&0.8112&28.93&0.8671&34.30&0.9494\\
			&&{\color{blue}34.86}&{\color{blue}0.9305}&{\color{blue}30.81}&{\color{blue}0.8505}&{\color{blue}29.40}&{\color{blue}0.8127}&{\color{blue}29.44}&{\color{blue}0.8767}&{\color{blue}34.95}&{\color{blue}0.9521}\\
			&&{\color{red}34.93}&{\color{red}0.9310}&{\color{red}30.88}&{\color{red}0.8514}&{\color{red}29.43}&{\color{red}0.8133}&{\color{red}29.57}&{\color{red}0.8784}&{\color{red}35.11}&{\color{red}0.9528}\\
			\hline\hline
			\multirow{11}{1.8cm}{Bicubic\newline SRCNN~\cite{dong2014learning}\newline VDSR~\cite{kim2016accurate}\newline LapSRN~\cite{lai2017deep}\newline MemNet~\cite{tai2017memnet}\newline EDSR~\cite{lim2017enhanced}\newline RDN~\cite{zhang2018residual}\newline RCAN~\cite{zhang2018image}\newline SAN~\cite{dai2019second}\newline ADCSR\newline ADCSR+}&\multirow{11}{*}{$\times4$}&28.42&0.8104&26.00&0.7027&25.96&0.6675&23.14&0.6577&24.89&0.7866 \\
			& &30.45&0.8628&27.50&0.7513&26.90&0.7101&24.52&0.7221&27.58&0.8555\\
			& &31.35&0.8830&28.02&0.7680&27.29&0.7251&25.18&0.7540&28.83&0.8870\\
			& &31.54&0.8850&28.19&0.7720&27.32&0.7270&25.21&0.7560&29.09&0.8900\\
			& &31.74&0.8893&28.26&0.7723&27.40&0.7281&25.50&0.7630&29.42&0.8942\\
			& &32.46&0.8968&28.80&0.7876&27.71&0.7420&26.64&0.8033&31.02&0.9148\\
			& &32.47&0.8990&28.81&0.7871&27.72&0.7419&26.61&0.8028&31.00&0.9173\\
			& &32.63&0.9002&28.87&0.7889&27.77&0.7436&26.82&0.8087&30.40&0.9082\\
			& &32.64&0.9003&28.92&0.7888&27.78&0.7436&26.79&0.8068&31.18&0.9169\\
			&&{\color{blue}32.77}&{\color{blue}0.9013}&{\color{blue}29.02}&{\color{blue}0.7917}&{\color{blue}27.86}&{\color{blue}0.7457}&{\color{blue}27.15}&{\color{blue}0.8174}&{\color{blue}31.76}&{\color{blue}0.9212}\\
			&&{\color{red}32.82}&{\color{red}0.9020}&{\color{red}29.09}&{\color{red}0.7930}&{\color{red}27.90}&{\color{red}0.7466}&{\color{red}27.27}&{\color{red}0.8197}&{\color{red}31.98}&{\color{red}0.9232}\\
			\hline
		\end{tabular}
	\end{center}
\end{table*}

\begin{figure*}[htp]
	\centering
	\includegraphics[scale=0.5]{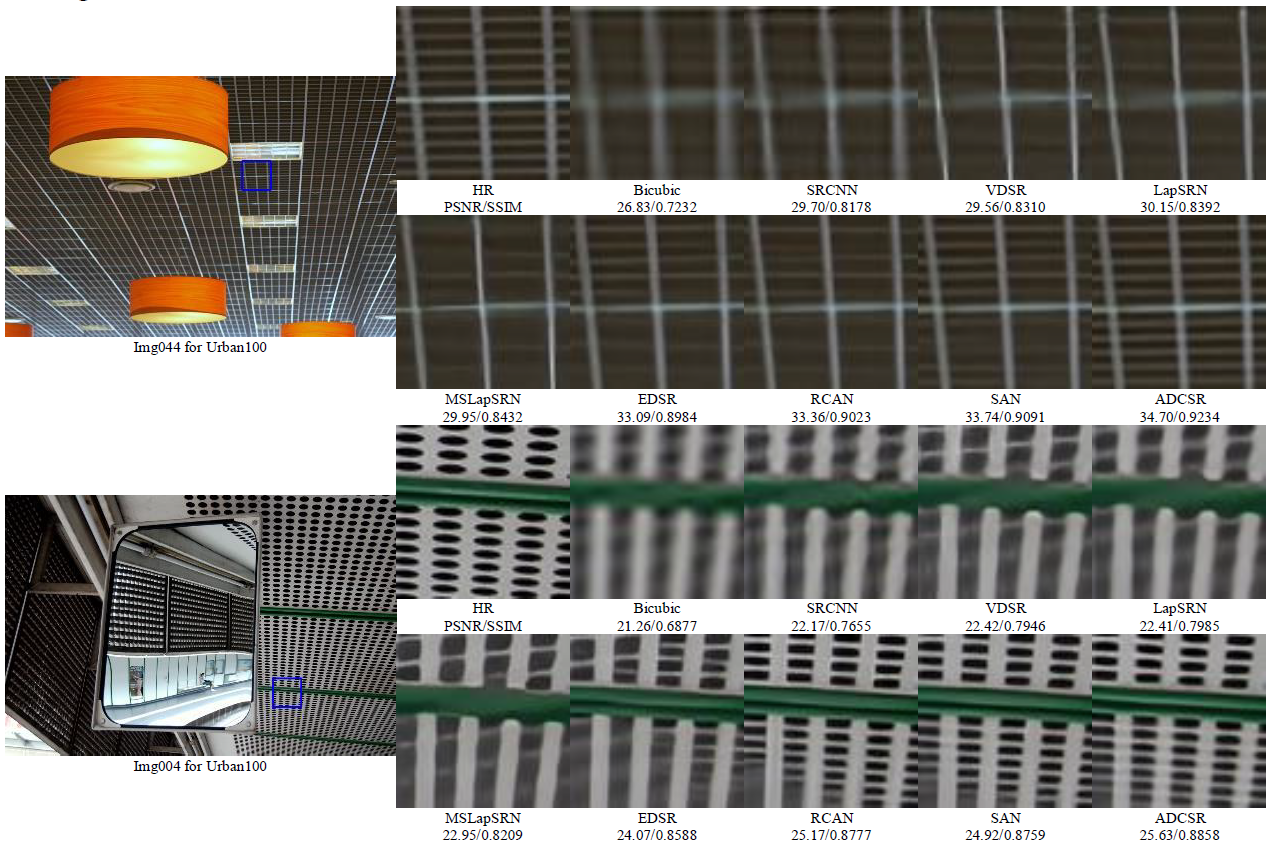}
	\caption{Visual results with bicubic degradation model($\times4$) on Urban100}
	\label{fig:cmpic}
\end{figure*}

\begin{figure*}[htp]
	\centering
	\includegraphics[scale=0.35]{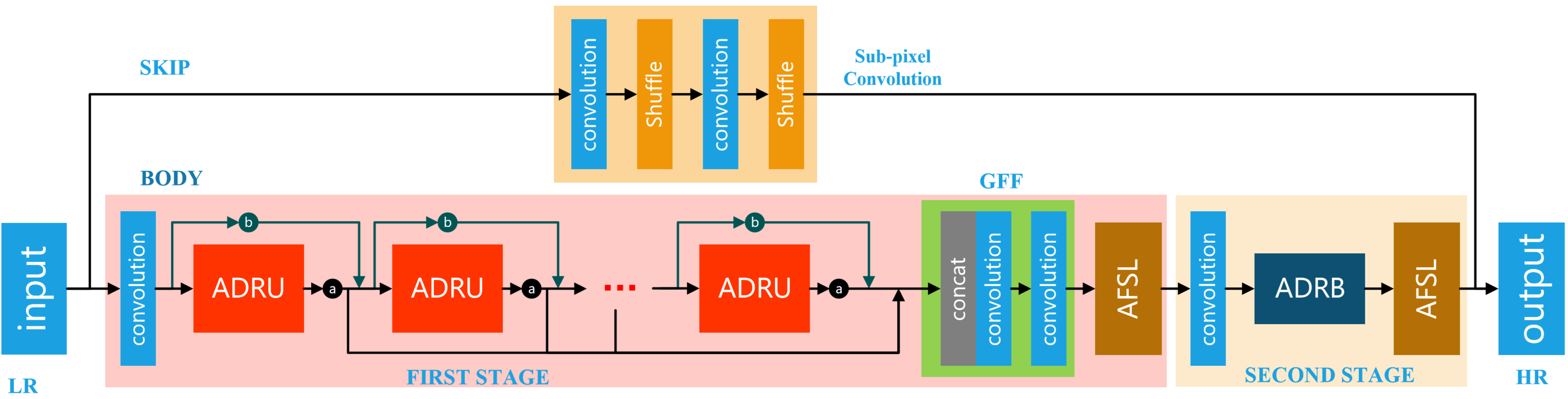}
	\caption{Two-stage adaptive dense connection super-resolution reconstruction network (DSSR)}
	\label{fig:dssr}
\end{figure*}

\subsection{Results with Bicubic Degradation}
In order to verify the validity of the model , we compare the performance on five standard benchmark datasets: Set5~\cite{bevilacqua2012low}, Set14~\cite{yang2010image}, B100~\cite{martin2001database}, Urban100~\cite{huang2015single}, and manga109~\cite{matsui2017sketch}. In terms of PSNR, SSIM and visual effects, We compare our models with the state-of-the-art methods including Bicubic, SRCNN~\cite{dong2014learning}, VDSR~\cite{kim2016accurate}, LapSRN~\cite{lai2017deep}, MemNet~\cite{tai2017memnet}, EDSR~\cite{lim2017enhanced}, RDN~\cite{zhang2018residual}, RCAN~\cite{zhang2018image}, SAN~\cite{dai2019second}. We also adopt self-ensemble strategy~\cite{lim2017enhanced}  to further improve our ADCSR and denote the self-ensembled ADCSR as ADCSR+. The results are shown in Table \ref{tab:qecm}. As can be seen from the table, the PSNR and SSIM of the algorithm in $\times2$, $\times3$, $\times4$ exceed the current state of the art.

Figure \ref{fig:cmpic} show the Qualitative comparison of our models with Bicubic, SRCNN~\cite{dong2014learning}, VDSR~\cite{kim2016accurate}, LapSRN~\cite{lai2017deep}, MSLapSRN~\cite{lai2018fast}, EDSR~\cite{lim2017enhanced}, RCAN~\cite{zhang2018image}, and SAN~\cite{dai2019second} . The images of SRCNN, EDSR, and RCAN are derived from the author's open-source model and code. Test images for VDSR, LapSRN, MSLapSRN, SAN are provided by their respective authors. In the comparison chart of img044 in Figure  \ref{fig:cmpic}, the image reconstructed by the algorithm is clear and close to the original image. In img004, our algorithm has a better visual effect.

\section{AIM2019: Extreme Super-Resolution Challenge}
This work is initially proposed for the purpose of participating in the AIM2019 Extreme Super-Resolution Challenge. The goal of the contest is to super-resolve an input image to an output image with a magnification factor $\times16$ and the challenge is called extreme super-resolution. 

Our model is the improved ADCSR, a two-stage adaptive dense connection super-resolution reconstruction network (DSSR). As shown in Figure \ref{fig:dssr}, the DSSR consists of two parts, SKIP and BODY. The SKIP is a simple sub-pixel convolution~\cite{shi2016real}. The BODY part is divided into two stages. The first stage includes a feature extraction layer, multiple ADRUs (adaptive, dense residual units), GFF (global feature fusion layer)~\cite{zhang2018residual}, and an AFSL layer (adaptive feature sub-pixel reconstruction layer).The second stage includes a feature amplification layer, an ADRB (adaptive dense residual block), and an AFSL.

During the training of DSSR, the network converges slowly due to the large network. We divide the network into two parts for training to speed up network convergence. When training DSSR, we first train the SKIP. The network ADCSR of scale is used as a pre-training parameter while training the entire network. At the same time, the feature extraction layer of the first level and each ADRU are set to be untrainable. During the period, GFF, AFSL and later second-level network parameters are trained at normal learning rates $1\times10^{-4}$. Finally, we train the entire network when the learning rate is small. We train DSSR with dataset DIV8K. Other training settings are the same as ADCSR.
Our model final result on the full resolution of the DIV8K test images is ($\times16$) :
PSNR = 26.79, SSIM = 0.7289.

\section{Conclusions}
We propose an adaptive densely connected super-resolution reconstruction algorithm (ADCSR). The algorithm is divided into two parts: BODY and SKIP. BODY improves the utilization of convolution features by adaptively dense connections. We also explore an adaptive sub-pixel reconstruction layer (AFSL) to reconstruct the features of the BODY output. We pre-train SKIP in advance so that the BODY focuses on high-frequency feature learning. Several comparative experiments demonstrate the effectiveness of the proposed improved method. On the standard datasets, the comparisons of PSNR, SSIM, and visual effects show that the proposed algorithm is superior to the state-of-the-art algorithms.

{\small
\bibliographystyle{ieee_fullname}
\bibliography{egbib}

\begin{thebibliography}{10}\itemsep=-1pt

\bibitem{bevilacqua2012low}
Marco Bevilacqua, Aline Roumy, Christine Guillemot, and Marie~Line
  Alberi-Morel.
\newblock Low-complexity single-image super-resolution based on nonnegative
  neighbor embedding.
\newblock 2012.

\bibitem{cao2019fast}
Yanpeng Cao, Zewei He, Zhangyu Ye, Xin Li, Yanlong Cao, and Jiangxin Yang.
\newblock Fast and accurate single image super-resolution via an energy-aware
  improved deep residual network.
\newblock {\em Signal Processing}, 162:115--125, 2019.

\bibitem{chang2004super}
Hong Chang, Dit-Yan Yeung, and Yimin Xiong.
\newblock Super-resolution through neighbor embedding.
\newblock In {\em Proceedings of the 2004 IEEE Computer Society Conference on
  Computer Vision and Pattern Recognition, 2004. CVPR 2004.}, volume~1, pages
  I--I. IEEE, 2004.

\bibitem{dai2019second}
Tao Dai, Jianrui Cai, Yongbing Zhang, Shu-Tao Xia, and Lei Zhang.
\newblock Second-order attention network for single image super-resolution.
\newblock In {\em Proceedings of the IEEE Conference on Computer Vision and
  Pattern Recognition}, pages 11065--11074, 2019.

\bibitem{dong2014learning}
Chao Dong, Chen~Change Loy, Kaiming He, and Xiaoou Tang.
\newblock Learning a deep convolutional network for image super-resolution.
\newblock In {\em European conference on computer vision}, pages 184--199.
  Springer, 2014.

\bibitem{he2016deep}
Kaiming He, Xiangyu Zhang, Shaoqing Ren, and Jian Sun.
\newblock Deep residual learning for image recognition.
\newblock In {\em Proceedings of the IEEE conference on computer vision and
  pattern recognition}, pages 770--778, 2016.

\bibitem{Huang_2017_CVPR}
Gao Huang, Zhuang Liu, Laurens van~der Maaten, and Kilian~Q. Weinberger.
\newblock Densely connected convolutional networks.
\newblock In {\em The IEEE Conference on Computer Vision and Pattern
  Recognition (CVPR)}, July 2017.

\bibitem{huang2015single}
Jia-Bin Huang, Abhishek Singh, and Narendra Ahuja.
\newblock Single image super-resolution from transformed self-exemplars.
\newblock In {\em Proceedings of the IEEE Conference on Computer Vision and
  Pattern Recognition}, pages 5197--5206, 2015.

\bibitem{karras2017progressive}
Tero Karras, Timo Aila, Samuli Laine, and Jaakko Lehtinen.
\newblock Progressive growing of gans for improved quality, stability, and
  variation.
\newblock {\em arXiv preprint arXiv:1710.10196}, 2017.

\bibitem{kim2016accurate}
Jiwon Kim, Jung Kwon~Lee, and Kyoung Mu~Lee.
\newblock Accurate image super-resolution using very deep convolutional
  networks.
\newblock In {\em Proceedings of the IEEE conference on computer vision and
  pattern recognition}, pages 1646--1654, 2016.

\bibitem{kim2016deeply}
Jiwon Kim, Jung Kwon~Lee, and Kyoung Mu~Lee.
\newblock Deeply-recursive convolutional network for image super-resolution.
\newblock In {\em Proceedings of the IEEE conference on computer vision and
  pattern recognition}, pages 1637--1645, 2016.

\bibitem{lai2017deep}
Wei-Sheng Lai, Jia-Bin Huang, Narendra Ahuja, and Ming-Hsuan Yang.
\newblock Deep laplacian pyramid networks for fast and accurate
  super-resolution.
\newblock In {\em Proceedings of the IEEE conference on computer vision and
  pattern recognition}, pages 624--632, 2017.

\bibitem{lai2018fast}
Wei-Sheng Lai, Jia-Bin Huang, Narendra Ahuja, and Ming-Hsuan Yang.
\newblock Fast and accurate image super-resolution with deep laplacian pyramid
  networks.
\newblock {\em IEEE transactions on pattern analysis and machine intelligence},
  2018.

\bibitem{ledig2017photo}
Christian Ledig, Lucas Theis, Ferenc Husz{\'a}r, Jose Caballero, Andrew
  Cunningham, Alejandro Acosta, Andrew Aitken, Alykhan Tejani, Johannes Totz,
  Zehan Wang, et~al.
\newblock Photo-realistic single image super-resolution using a generative
  adversarial network.
\newblock In {\em Proceedings of the IEEE conference on computer vision and
  pattern recognition}, pages 4681--4690, 2017.

\bibitem{li2019feedback}
Zhen Li, Jinglei Yang, Zheng Liu, Xiaomin Yang, Gwanggil Jeon, and Wei Wu.
\newblock Feedback network for image super-resolution.
\newblock In {\em Proceedings of the IEEE Conference on Computer Vision and
  Pattern Recognition}, pages 3867--3876, 2019.

\bibitem{lim2017enhanced}
Bee Lim, Sanghyun Son, Heewon Kim, Seungjun Nah, and Kyoung Mu~Lee.
\newblock Enhanced deep residual networks for single image super-resolution.
\newblock In {\em Proceedings of the IEEE conference on computer vision and
  pattern recognition workshops}, pages 136--144, 2017.

\bibitem{martin2001database}
David Martin, Charless Fowlkes, Doron Tal, Jitendra Malik, et~al.
\newblock A database of human segmented natural images and its application to
  evaluating segmentation algorithms and measuring ecological statistics.
\newblock Iccv Vancouver:, 2001.

\bibitem{matsui2017sketch}
Yusuke Matsui, Kota Ito, Yuji Aramaki, Azuma Fujimoto, Toru Ogawa, Toshihiko
  Yamasaki, and Kiyoharu Aizawa.
\newblock Sketch-based manga retrieval using manga109 dataset.
\newblock {\em Multimedia Tools and Applications}, 76(20):21811--21838, 2017.

\bibitem{schulter2015fast}
Samuel Schulter, Christian Leistner, and Horst Bischof.
\newblock Fast and accurate image upscaling with super-resolution forests.
\newblock In {\em Proceedings of the IEEE Conference on Computer Vision and
  Pattern Recognition}, pages 3791--3799, 2015.

\bibitem{shi2016real}
Wenzhe Shi, Jose Caballero, Ferenc Husz{\'a}r, Johannes Totz, Andrew~P Aitken,
  Rob Bishop, Daniel Rueckert, and Zehan Wang.
\newblock Real-time single image and video super-resolution using an efficient
  sub-pixel convolutional neural network.
\newblock In {\em Proceedings of the IEEE conference on computer vision and
  pattern recognition}, pages 1874--1883, 2016.

\bibitem{shi2013cardiac}
Wenzhe Shi, Jose Caballero, Christian Ledig, Xiahai Zhuang, Wenjia Bai, Kanwal
  Bhatia, Antonio M Simoes~Monteiro de Marvao, Tim Dawes, Declan O’Regan, and
  Daniel Rueckert.
\newblock Cardiac image super-resolution with global correspondence using
  multi-atlas patchmatch.
\newblock In {\em International Conference on Medical Image Computing and
  Computer-Assisted Intervention}, pages 9--16. Springer, 2013.

\bibitem{tai2017memnet}
Ying Tai, Jian Yang, Xiaoming Liu, and Chunyan Xu.
\newblock Memnet: A persistent memory network for image restoration.
\newblock In {\em Proceedings of the IEEE international conference on computer
  vision}, pages 4539--4547, 2017.

\bibitem{timofte2013anchored}
Radu Timofte, Vincent De~Smet, and Luc Van~Gool.
\newblock Anchored neighborhood regression for fast example-based
  super-resolution.
\newblock In {\em Proceedings of the IEEE international conference on computer
  vision}, pages 1920--1927, 2013.

\bibitem{tong2017image}
Tong Tong, Gen Li, Xiejie Liu, and Qinquan Gao.
\newblock Image super-resolution using dense skip connections.
\newblock In {\em Proceedings of the IEEE International Conference on Computer
  Vision}, pages 4799--4807, 2017.

\bibitem{wang2019lightweight}
Chaofeng Wang, Zheng Li, and Jun Shi.
\newblock Lightweight image super-resolution with adaptive weighted learning
  network.
\newblock {\em arXiv preprint arXiv:1904.02358}, 2019.

\bibitem{xu2019towards}
Xiangyu Xu, Yongrui Ma, and Wenxiu Sun.
\newblock Towards real scene super-resolution with raw images.
\newblock In {\em Proceedings of the IEEE Conference on Computer Vision and
  Pattern Recognition}, pages 1723--1731, 2019.

\bibitem{yang2010image}
Jianchao Yang, John Wright, Thomas~S Huang, and Yi Ma.
\newblock Image super-resolution via sparse representation.
\newblock {\em IEEE transactions on image processing}, 19(11):2861--2873, 2010.

\bibitem{yu2018wide}
Jiahui Yu, Yuchen Fan, Jianchao Yang, Ning Xu, Zhaowen Wang, Xinchao Wang, and
  Thomas Huang.
\newblock Wide activation for efficient and accurate image super-resolution.
\newblock {\em arXiv preprint arXiv:1808.08718}, 2018.

\bibitem{zhang2019deep}
Kai Zhang, Wangmeng Zuo, and Lei Zhang.
\newblock Deep plug-and-play super-resolution for arbitrary blur kernels.
\newblock In {\em Proceedings of the IEEE Conference on Computer Vision and
  Pattern Recognition}, pages 1671--1681, 2019.

\bibitem{zhang2019zoom}
Xuaner Zhang, Qifeng Chen, Ren Ng, and Vladlen Koltun.
\newblock Zoom to learn, learn to zoom.
\newblock In {\em Proceedings of the IEEE Conference on Computer Vision and
  Pattern Recognition}, pages 3762--3770, 2019.

\bibitem{zhang2018image}
Yulun Zhang, Kunpeng Li, Kai Li, Lichen Wang, Bineng Zhong, and Yun Fu.
\newblock Image super-resolution using very deep residual channel attention
  networks.
\newblock In {\em Proceedings of the European Conference on Computer Vision
  (ECCV)}, pages 286--301, 2018.

\bibitem{zhang2018residual}
Yulun Zhang, Yapeng Tian, Yu Kong, Bineng Zhong, and Yun Fu.
\newblock Residual dense network for image super-resolution.
\newblock In {\em Proceedings of the IEEE Conference on Computer Vision and
  Pattern Recognition}, pages 2472--2481, 2018.

\bibitem{zou2011very}
Wilman~WW Zou and Pong~C Yuen.
\newblock Very low resolution face recognition problem.
\newblock {\em IEEE Transactions on image processing}, 21(1):327--340, 2011.

\end{thebibliography}
}

\end{document}